\documentclass[12pt]{spieman}

\usepackage{amsmath,amsfonts,amssymb}
\usepackage{graphicx}
\usepackage{setspace}
\usepackage{tocloft}
\usepackage{comment}
\usepackage{subcaption}
\usepackage[utf8]{inputenc}

\usepackage[numbers,sort&compress]{natbib}

\title{Enhancing Imaging Depth and Sensitivity in Reflectance Mode Near Infrared Optical Imaging with Scatter Reducing Agents
}

\author[a]{Mannu Bardhan Paul}
\author[b]{Kaiser Niknam}
\author[b,a,c,*]{Mini Das}

\affil[a]{Department of Biomedical Engineering, University of Houston, Houston, TX, 77204, USA}
\affil[b]{Department of Physics, University of Houston, Houston, TX, 77204, USA}
\affil[c]{Department of Electrical and Computer Engineering, University of Houston, Houston, TX, 77204}

\cftpagenumbersoff{figure}
\cftpagenumbersoff{table}

\begin{document}
\maketitle

\begin{abstract}
Continuous-wave (CW) near-infrared (NIR) diffuse optical measurements are widely used due to their simplicity and suitability for portable applications, but their sensitivity to deep tissue is fundamentally limited by strong scattering. In this study, we investigate the role of scatter reducing agents in a reflectance mode imaging setting. We use food-grade dye Tartrazine as a scatter reducing agent to enhance depth sensitivity and weak-absorber detectability in CW diffuse reflectance measurements. Induced transparency and scatter reduction was characterized through spectral and refractive-index measurements, confirming reduced scattering in the near-infrared range. Prior work showed improvement in transmission effects and imaging using this scatter reduction mechanism. However, benefits in reflectance mode has not been examined. We found that reflectance signal was enhanced when the dye was applied on chicken breast phantom. However, we saw reduced reflectance sensitivity when the dye was uniformly dissolved in intralipid phantom which is a commonly used for NIR imaging studies. This shows that the gradient of  refractive index modulation created as the dye diffuses from the top layer allows increased reflectance signal sensitivity of optical photons. However, when the scatter reduction is uniform throughout the phantom (like in intralipid phantom), the improved reflectance sensitivity was not observed. Our study points to significant redistribution of photons with scatter modulation with Tartrazine dye. We show significant improvement in sensitivity to signals with reflectance imaging. In this work, we also explore physics models and Monte Carlo simulations  To elucidate the underlying mechanism of dye induced scatter reduction in tissue, analytical diffusion models and Monte Carlo simulations were employed. Modeling results show the impact of refractive index gradient created due to dye diffusion in enhancing reflectance sensitivity. These findings demonstrate that dye induced scatter reduction provides a practical, low-complexity approach to improving depth sensitivity in CW diffuse reflectance measurements and extend the functional capabilities of CW-NIRS systems for deep-tissue sensing applications. Our preliminary studies shows up to five fold enhancement in signal sensitivity for signals between two and three cm depth.
\end{abstract}


{\noindent \footnotesize\textbf{*}\linkable{mdas@uh.edu} }

\begin{spacing}{1.2}   

\section{Introduction}
\label{sect:intro}  

Near-infrared (NIR) diffuse optical imaging (DOI) is widely used in clinical and research settings for neurovascular and hemodynamic study based on its sensitivity to properties like hemoglobin concentration, oxygenation, blood volume  \citep{Jones2016,Ghanayem2016}. Beyond neuroscience, NIR optical techniques have also been explored for applications such as tissue characterization and monitoring \citep{hamaoka2007near,ferrari2012brief}. Its noninvasive and ionizing-radiation-free nature enables safe, repeated, and longitudinal measurements, including in clinically vulnerable patient populations. 

Despite these advantages, deep-tissue DOI is fundamentally constrained by strong multiple scattering in biological tissue. Light attenuation is governed by both absorption ($\mu_a$) and scattering ($\mu_s$) coefficients; however, in the NIR spectral range, scattering typically exceeds absorption by one to two orders of magnitude \citep{Jacques2013,Martelli2022,Taroni2003}. Consequently, photons undergo many scattering events and propagate diffusively, causing detected reflectance signals to be dominated by superficial tissue contributions and reducing sensitivity to deeper regions of interest \citep{Jacques2013,Martelli2012}, particularly for the weak absorbing signals.

Diffuse optical imaging systems are commonly implemented using continuous-wave (CW), frequency-domain (FD), or time-domain (TD) detection schemes. FD and TD approaches provide additional measurement dimensions (e.g., phase or time-of-flight) that can improve detection enhancing depth discrimination \citep{Fantini2012,Torricelli2014}. However, these systems require increased hardware complexity, calibration effort, and cost; limiting practicality for portable, point-of-care, or longitudinal use. In contrast, CW systems measure intensity changes arising from absorption variations, enabling robust and low-cost instrumentation. Measurements solely based on intensity features lack depth sensitivity, and many CW studies operate within effective sensitivity depths of only a few centimeters \citep{Grosenick2016,Erickson2009}. Despite these constraints, CW systems remain widely used due to their simplicity and practicality \citep{Grosenick2016,Choe2009}.

These depth-sensitivity limitations are particularly relevant in applications such as breast DOI, where weak subsurface physiological contrast is diluted with high tissue scattering. Breast cancer remains one of the most prevalent malignancies worldwide, and early detection is strongly associated with improved outcomes \citep{Zhang2025,Grosenick2016}. Optical techniques have been explored as complementary tools for probing breast tissue physiology, including vascular remodeling, angiogenesis, and oxygenation-related contrast \citep{Eelen2020,Bisht2021}. In current diagnostic protocols, X-ray mammography remains the clinical standard, but sensitivity is reduced in dense breast tissue \citep{Mota2025,Taylor2024}, and ionizing radiation limits frequent monitoring. Ultrasound and magnetic resonance imaging provide additional structural and functional information but involve trade-offs in cost, accessibility, spatial resolution, or operator dependence \citep{Gilbert2019,Carp2013,Raikhlin2015}. Here, breast imaging is used as a motivating example of a clinically relevant scenario where scattering-driven superficial weighting limits the detectability of weak subsurface contrast in CW diffuse optical measurements.

Several strategies have been proposed to mitigate the dominance of superficial sensitivity in CW DOI, including optimized source--detector geometries, multi-distance measurements, and algorithmic approaches designed to suppress superficial contributions \citep{Brigadoi2015,gagnon2012short}. While these methods can improve depth sensitivity to some extent, they do not fundamentally alter the physical dominance of scattering in superficial layers. This motivates approaches that directly modify tissue optical properties to alter photon transport pathways. Optical clearing can reduce scattering by matching refractive-indices among tissue constituents. Classical clearing approaches often rely on dehydration, delipidification, or structural reorganization, which are difficult to achieve and maintain in living tissues \citep{richardsonClarifyingTissueClearing2015b,hrTissueClearingIts2020}. A number of studies have been done focusing on thin tissues and microscopy- or transmission-based imaging modalities \citep{ZhuOCA,Genina2010,Genina2022,Ziaee:25}. In exogenous agent–based optical clearing, scattering is reduced by elevating the refractive index of the interstitial medium, thereby reducing refractive-index mismatches at cellular and subcellular interfaces while leaving tissue microstructure largely intact \citep{Tuchin1997,Tuchin2005}. This approach has been extensively studied in terms of agent efficacy, tissue permeability, and depth-dependent clearing dynamics \citep{Genina2010,Oliveira2019,Tainaka2018,Tuchin2005}.

Recently, the food dye Tartrazine (FD\&C Yellow~5) has been identified as an efficient optical clearing agent in the red to near-infrared spectral region. At sufficiently high concentrations, Tartrazine produces a large increase in the refractive index of aqueous media and has been reported to reduce tissue scattering more effectively than conventional agents such as glycerol \citep{Ou2024}. Recent modeling studies, including work by Fu \textit{et al.} \citep{Fu2025}, further suggest that reducing superficial scattering through refractive-index modulation can enhance photon penetration and absorption in deeper tissue layers. However, the implications of such agent-based optical clearing for weak-absorber detectability in CW diffuse reflectance measurements at large source--detector separations remain underexplored.

In this study, we investigate how superficial scattering reduction using Tartrazine alters photon transport and improves weak-signal detectability at depth in CW diffuse reflectance measurements. Using controlled phantom experiments, we investigate how optical clearing improves weak-absorber detectability in CW diffuse reflectance measurements. Experimental observations are supported and validated by both analytical models and Monte Carlo simulations, showing that the improvement arises from deeper photon penetration and increased absorption at depth, rather than from increased photon throughput.

\section{Methods and Materials}

\subsection{Experimental purpose and methodological flow}

\begin{figure}[t]
  \centering
  \includegraphics[width=0.8\linewidth]{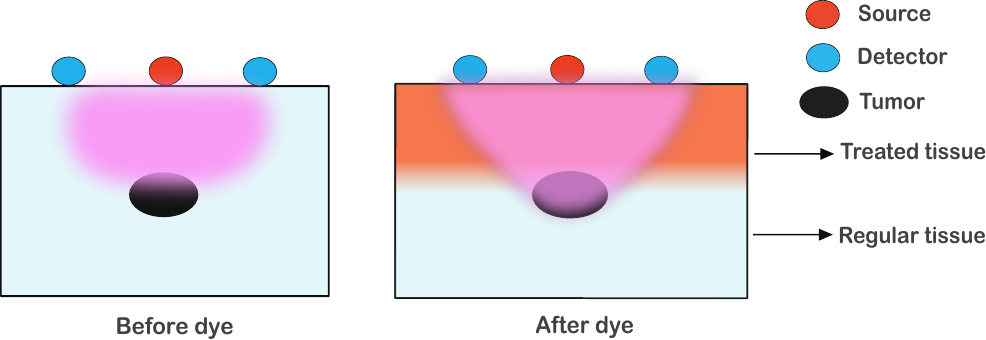}
  \caption{Conceptual illustration of Tartrazine-induced optical clearing. Topical application reduces superficial scattering and redistributes photons toward deeper tissue represented by the pink region, enhancing subsurface signal detection.}
  \label{fig:SignalWIthdye}
\end{figure}

The methodology of this study was designed to examine how selective reduction of superficial tissue scattering influences subsurface signal detectability in CW diffuse reflectance measurements. The food-grade dye Tartrazine was used as an optical clearing agent due to its ability to reduce scattering through refractive-index modulation without inducing structural changes in tissue. Fresh chicken muscle tissue was selected as the base phantom because it exhibits optical scattering properties in the NIR region that are comparable to those of human soft tissue \citep{Cheong1990} and provides sufficient mechanical integrity to support controlled, depth-dependent diffusion of the clearing agent. When applied topically, Tartrazine diffuses preferentially into the superficial tissue region, producing a depth-dependent modification of optical properties. This behavior enables controlled investigation of how suppression of superficial scattering alters photon transport in reflection-mode geometry by reducing early backscattering and redistributing photons toward deeper tissue regions, thereby enhancing interaction with subsurface absorbers, as illustrated conceptually in Fig.~\ref{fig:SignalWIthdye}. To distinguish the effects of layered versus homogeneous scattering modification, a uniform liquid phantom was also examined as a comparative control. CW measurements were performed at large source–detector separations, where sensitivity to deep tissue absorbers is typically limited. Signal detectability was quantified by measuring intensity changes in response to controlled subsurface absorption perturbations introduced via ink injection.

\subsection{Mechanism of optical clearing with agent-based refractive-index modulation}

Biological tissue is an optically heterogeneous medium composed of cellular and subcellular constituents with distinct refractive indices (RI). Microscopic RI mismatches between aqueous compartments and lipid- or protein-rich structures give rise to strong elastic scattering, which dominates light transport in the visible and NIR spectral ranges. Optical clearing agents (OCAs) reduce tissue scattering by increasing the RI of low-index aqueous regions, thereby decreasing RI mismatch with higher-index structural components.

In soft tissues, aqueous components such as cytosol and interstitial fluid typically exhibit RI values of approximately 1.35–1.37, whereas lipid- and protein-rich structures, including cell membranes, collagen fibrils, and organelles, possess higher RIs in the range of 1.39–1.47 \citep{Tuchin2015}. Because scattering strength depends strongly on this mismatch, even modest increases in the RI of the interstitial medium can substantially reduce scattering. Both semi-empirical scattering models and experimental studies have shown that increasing the RI of the interstitial medium suppresses scattering without altering tissue geometry \citep{Graaff1992,Tuchin1997}. When applied topically, optical clearing agents produce this effect through diffusion into tissue, with penetration depth governed by tissue permeability, microstructure, and molecular diffusivity \citep{Genina2022}. Under commonly used one-dimensional diffusion assumptions, the agent concentration follows Fick’s law , producing a depth-dependent profile characterized by a diffusion length scale $\delta \sim \sqrt{Dt}$, where $D$ is the effective diffusion coefficient and $t$ is the application time \citep{kotykMembraneTransportInterdisciplinary1977}. For experimental time scales on the order of tens of minutes, OCAs penetrate only a few millimeters into soft tissue, leading to a localized reduction in the reduced scattering coefficient \citep{Tuchin1997}.

Within this framework, the local refractive index of the interstitial medium can be expressed as $n(z,t) = n_0 + \Delta n[C(z,t)]$, where $C(z,t)$ denotes the concentration of the optical clearing agent. Under commonly used microstructural scattering approximations, variations in tissue scattering are dominated by changes in refractive-index mismatch between scattering structures and the surrounding medium. Following Tuchin \citep{Tuchin1997}, this effect can be described in terms of the refractive-index ratio $m = n_s / n_I$, where $n_s$ is the refractive index of the scattering structures and $n_I$ is that of the interstitial medium.

For two different degrees of refractive-index matching, before clearing ($m_1 = n_s / n_0$) and after clearing ($m_2 = n_s / n(z,t)$), the scattering coefficients are
approximately related by \citep{Tuchin1997} :
\begin{equation}
\frac{\mu_s(z,t)}{\mu_{s0}} \approx
\left( \frac{m_2^2 - 1}{m_1^2 - 1} \right)^2 ,
\label{eq:mus_tuchin}
\end{equation}
where $\mu_{s0}$ and $\mu_s(z,t)$ denote the scattering coefficients before and after optical clearing, respectively. As the clearing agent elevates the RI of the interstitial medium $n(z,t)$, the mismatch between $n_I$ and the scatterer refractive index $n_s$ is reduced, leading to a strong suppression of elastic scattering. Consequently, even modest increases can result in substantial reductions in $\mu_s$.

In the muscle tissue phantom, topical OCA application creates a depth-dependent scattering profile, reducing $\mu_s$ near the surface while leaving deeper regions largely unchanged. This depth-dependent scattering modification motivates the use of layered diffusion theory and Monte Carlo simulations in subsequent sections to examine how it alters photon transport and CW diffuse reflectance at large source–detector separations.

\subsection{System development and measurement principle}

Diffuse reflectance measurements were performed using a  CW-fNIRS system (NIRSport 2, NIRx Medical Technologies LLC). The system is equipped with dual-wavelength light sources at 760 and 850 nm and comprises 8 sources and 8 detectors, with a maximum source output power of 32 mW. Its modular architecture supports flexible optode configurations and up to 256 measurement channels depending on probe geometry. All source and detector optodes were positioned on the top surface of the tissue phantom using standard elastic caps. Optode locations were digitized in three dimensions to ensure accurate spatial registration. The system records voltage signals proportional to detected light intensity, which constitute the raw measurement output. All measurements were conducted following the manufacturer’s recommended operating procedures.

\begin{figure}[t]
\centering
\includegraphics[width=1\linewidth]{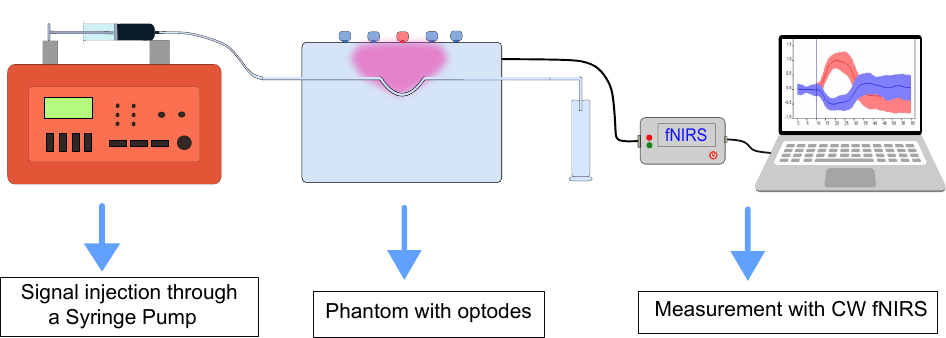}
\caption{Schematic of the experimental setup for CW diffuse reflectance measurements. A syringe pump delivers controlled ink injections to create a subsurface absorbing inclusion within the tissue phantom. Optodes are placed on the phantom surface to record diffuse reflectance intensity before, during, and after ink infusion, enabling assessment of depth-dependent signal detectability under varying optical clearing conditions.}
\label{fig:Schematic_syringepump}
\end{figure}

To enable time-dependent contrast measurements using a CW system, controlled changes in optical absorption were introduced via ink injection, as illustrated in Fig.~\ref{fig:Schematic_syringepump}. During each experiment, the CW-fNIRS system continuously recorded diffuse reflectance intensity before, during, and after ink infusion. The absorbing inclusion was positioned at an approximate depth of 2.5 cm within the phantom. Because the region of maximum sensitivity in diffuse reflectance measurements occurs at roughly half the source–detector separation, measurements acquired at a 6 cm source–detector distance were analyzed to maximize sensitivity to the subsurface absorber \citep{hanOptimizationSourcedetectorSeparation2021}.

In CW measurements, intensity changes arising from absorption variations are commonly described by the modified Beer–Lambert law \citep{Scholkmann2013}:
\begin{equation}
\Delta A = \log\left(\frac{I_b}{I_m}\right)
= \varepsilon(\lambda)\ \cdot\Delta C \cdot \mathrm{DPF}(d,\lambda)\cdot d,
\label{eq:MBLL}
\end{equation}

where $I_b$ denotes the baseline detected intensity measured in the absence of the subsurface absorber, i.e., when only the background medium is present, and $I_m$ denotes the detected intensity measured after introduction of the absorbing inclusion. $\varepsilon(\lambda)$ is the wavelength-dependent molar extinction coefficient of the absorber, $\Delta C$ represents the change in absorber concentration relative to baseline, and $\mathrm{DPF}(d,\lambda)$ is the differential pathlength factor accounting for increased photon pathlength due to multiple scattering at source--detector separation $d$.

Equation~(\ref{eq:MBLL}) is included to illustrate the theoretical relationship between absorption changes and detected intensity in CW systems. Concentration recovery is not performed in this study. All results therefore focus on relative intensity (voltage) changes and their redistribution under different scattering conditions, evaluated using tissue-mimicking phantoms with controlled absorber placement.

\subsubsection{Preparation of phantoms}

Fresh boneless chicken muscle tissue, purchased from a local supplier, was used as a tissue surrogate to model breast tissue optical behavior. Relatively homogeneous muscle regions were selected, and tissue blocks of sufficient size were prepared to accommodate multiple source–detector optode placements. Shallow surface indentations were created to seat optodes reproducibly and minimize variability in source–detector separation across repeated measurements. To introduce a controlled subsurface absorber, a narrow channel was created using a small-diameter drill, into which a transparent silicone rubber tube (0.3~mm inner diameter) was inserted at a predefined depth. This tubing served as a conduit for repeatable delivery of absorbing ink solutions without disturbing the surrounding tissue structure. Subsurface absorbers were introduced using Indian ink (Dr.\ Ph.\ Martin’s) diluted in water at different volume concentrations to mimic weak to strong absorbers that generate optical contrast in continuous-wave measurements \citep{Ninni2010}.

Optical clearing of the tissue surface was achieved by applying Tartrazine dye solutions prepared following established protocols reported by Ou \textit{et al.} \citep{Ou2024}. A dye concentration of 0.6~M was selected, as the prior study showed this concentration to produce most-efficient scattering reduction. Dye was applied in a layer-by-layer manner to systematically modulate superficial scattering while preserving the optical properties of deeper regions. One layer of dye application was defined as uniformly applying the Tartrazine solution to the tissue surface, followed by a 10-minute equilibration period prior to measurement. Additional layers were applied by repeating this procedure.

To provide a homogeneous-scattering control and to contrast layered versus uniform optical property modifications, a liquid tissue-mimicking phantom was also prepared. The phantom was contained in a bucket comparable in size to the chicken tissue block, with sources and detectors placed on the top surface and a subsurface tube positioned at a predefined depth in a similar geometry. Optical scattering properties comparable to soft tissue were achieved using a 20\% intralipid emulsion (Sigma-Aldrich, Inc., MO, USA) \citep{pogueReviewTissueSimulating2006,laiDependenceOpticalScattering2014}. For these control measurements, Tartrazine was uniformly mixed into the liquid phantom at equivalent concentrations for each measurement, producing a uniform modification of scattering throughout it.

In addition to the primary tissue phantoms, preliminary experiments were conducted using a range of tissue-mimicking materials. It includes both synthetic and organic compounds like PDMS, sodium polyacrylate, agar, and gelatin matrices; combined with scatterers such as SiO$_2$, TiO$_2$, milk, and intralipid \citep{pogueReviewTissueSimulating2006,Ayers2008,Ley2014}. These phantoms were evaluated to assess their ability to support depth-dependent optical clearing profiles under topical agent application. However, most liquid and semi-solid phantoms failed to maintain stable layered optical property gradients; as rapid mixing, diffusion, or structural relaxation led to spatially uniform scattering changes. This behavior contrasts with biological tissue, where microstructural heterogeneity and limited permeability naturally constrain agent diffusion, enabling sustained superficial clearing layers.

To independently characterize the refractive-index–mediated scattering reduction induced by Tartrazine, preliminary optical measurements were performed using SiO$_2$ particles (1--5~$\mu$m) as a controlled scattering medium. Its provides a well-defined and reproducible scattering environment in which refractive-index modulation can be isolated from biological variability. Spectral transmittance and absorption measurements were obtained using a UV–vis–NIR spectrophotometer (Shimadzu UV-1800, 200--1100~nm), and refractive-index dispersion was derived from the absorption spectra using the Kramers--Kronig relations \citep{saiDesigningRefractiveIndex2020b, rochaDeterminationOpticalConstants2014}. These measurements served as foundational validation of Tartrazine-induced refractive-index modulation and are presented in the Results section.

\subsection{Diffusion-approximation models for diffuse reflectance}

Photon propagation in highly scattering biological tissue is fundamentally governed by the radiative transfer equation (RTE) \citep{Ishimaru1978}. In the NIR spectral range, where reduced scattering dominates absorption ($\mu_s' \gg \mu_a$), photon transport becomes diffuse beyond a few transport mean free paths. Under these conditions, the RTE can be accurately approximated by the diffusion equation, providing an analytically tractable framework for modeling CW diffuse reflectance in homogeneous and layered tissues \citep{Patterson1989,Martelli2012,dasAnalyticalSolutionLight2006}.

Within the diffusion approximation, photon attenuation is characterized by the effective attenuation coefficient
\begin{equation}
\mu_{\mathrm{eff}} = \sqrt{3\mu_a(\mu_a + \mu_s')},
\end{equation}
which governs the spatial decay of fluence and diffuse reflectance. Because $\mu_s' \gg \mu_a$ in most soft tissues, $\mu_{\mathrm{eff}}$ is primarily controlled by the reduced scattering coefficient. As a result, changes in $\mu_s'$, particularly within superficial tissue layers, can strongly influence photon penetration depth and the spatial distribution of detected reflectance. This sensitivity motivates the use of diffusion-based models to examine how superficial scattering reduction alters CW diffuse reflectance.

Farrell \textit{et al.} derived an analytical solution for steady-state diffuse reflectance from a semi-infinite homogeneous medium using a diffusion dipole approximation \citep{Farrell1992}. In this formulation, the radial dependence of diffuse reflectance follows
\begin{equation}
R(r) \propto \frac{\exp(-\mu_{\mathrm{eff}} r)}{r},
\end{equation}
where $\mu_{\mathrm{eff}}=\sqrt{3\mu_a(\mu_a+\mu_s')}$ governs the radial decay rate. 
To apply this homogeneous model to our layered phantom, the depth-dependent optical properties are translated into an effective reduced scattering coefficient $\mu_{s,\mathrm{eff}}'$ using photon pathlength weights from the Monte Carlo model, and then evaluated the Farrell dipole solution using $(\mu_a,\mu_{s,\mathrm{eff}}')$. 
Here, optical clearing is represented by a decrease in $\mu_{s,\mathrm{eff}}'$, highlighting the central role of $\mu_{\mathrm{eff}}$ in radial decay, leading to increased predicted reflectance at larger source--detector separations.

Biological tissues, however, are inherently layered, with optical properties that vary with depth. To account for this structure, Schmitt and Kumar extended the diffusion framework to layered media, expressing diffuse reflectance as a superposition of contributions from individual layers \citep{Schmitt1990}:
\begin{equation}
R(r) \approx \frac{1}{r}\sum_{i=1}^{N} C_i \exp(-\mu_{\mathrm{eff},i} r),
\end{equation}
where $\mu_{\mathrm{eff},i}$ and $C_i$ denote the effective attenuation coefficient and weighting factor of the $i$th layer, respectively. In this model, optical clearing is incorporated by assigning reduced $\mu_s'$ values to superficial layers while maintaining baseline scattering in deeper regions. Layers with lower $\mu_{\mathrm{eff}}$ therefore contribute disproportionately to reflectance at larger radial distances.

Together, these diffusion models predict that reducing superficial scattering suppresses early photon randomization and near-source backscattering, while increasing the relative contribution of longer pathlength trajectories that exit the tissue at larger source–detector separations. This redistribution results in reduced reflectance near the source and enhanced signal at larger separations, a regime directly relevant to improving subsurface absorber detectability in CW measurements. These analytical predictions provide the physical basis for our hypothesis that optical clearing enhances depth sensitivity by redistributing photon transport rather than increasing total photon throughput. In the following sections, these predictions are examined using Monte Carlo simulations and validated experimentally, with Monte Carlo modeling employed to capture layered geometries and transport behavior beyond diffusion assumptions.

\subsection{Monte Carlo simulation setup}

To complement the analytical diffusion models and capture photon transport beyond the validity of the diffusion approximation, Monte Carlo (MC) simulations were performed using MCmatlab \citep{Marti_2018}. MC simulations explicitly track photon trajectories, scattering events, and absorption interactions in layered, highly scattering media; enabling direct evaluation of how superficial scattering reduction alters diffuse reflectance and subsurface absorption sensitivity.

Photon transport was simulated within a three-dimensional tissue volume of size
$15 \times 15 \times 5~\mathrm{cm}^3$ ($x \times y \times z$). A continuous-wave pencil-beam source was positioned at the center of the top surface ($z=0$), and diffuse reflectance was recorded at the same surface. For each simulation condition, $10^6$ photons were launched to ensure statistical convergence of reflectance and absorption metrics.

A subsurface absorbing target was placed at a depth of $1~\mathrm{cm}$ below the surface, corresponding to the ink-filled tube used in the experiments. The background optical properties were chosen to represent breast or muscle tissue in the near-infrared range, with absorption coefficient $\mu_a = 0.05~\mathrm{cm}^{-1}$ and anisotropy factor $g = 0.9$. Scattering coefficients were selected within the physiologically relevant range $\mu_s = 20$--$400~\mathrm{cm}^{-1}$, corresponding to a baseline scattering coefficient $\mu_s \approx 30~\mathrm{cm}^{-1}$, consistent with reported soft-tissue values \citep{Martelli2012,cubeddu1997solid,Pifferi2004,Durduran2002,Torricelli2014,Taroni2003}.

\begin{figure}[t]
  \centering
  \includegraphics[width=1.0\linewidth]{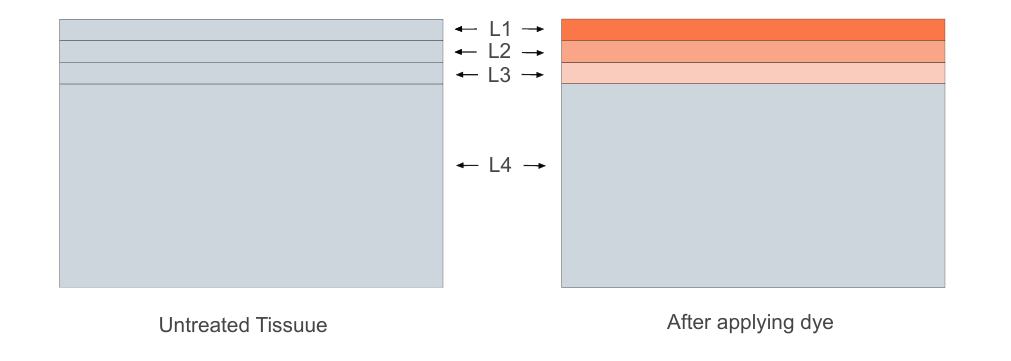}
 \caption{Layered tissue model used in Monte Carlo simulations before and after optical clearing.}

  \label{fig:Tissue_model_dye}
\end{figure}

Schematic illustration is presented in fig \ref{fig:Tissue_model_dye} of the layered tissue model used in Monte Carlo simulations. In untreated tissue (left), optical properties are assumed homogeneous throughout the volume. After topical application of Tartrazine (right), optical clearing produces a depth-dependent scattering profile consisting of multiple superficial layers (L1–L3) with progressively reduced scattering coefficients, while the deeper region (L4) retains baseline tissue properties. 
This layered configuration is modeled by selectively reducing the scattering coefficient in superficial tissue layers while leaving deeper regions, governed by equation \ref{eq:mus_tuchin}. Three scattering configurations were considered. In the no-dye condition, the tissue was assumed homogeneous with $\mu_s = 30~\mathrm{cm}^{-1}$ throughout the volume. In the one-layer clearing condition, the top 3 layers were assigned with $\mu_s = 5,10,20~\mathrm{cm}^{-1}$,respectively; while the remaining layer retained the baseline value. In the two-layer clearing condition, stronger superficial clearing was modeled by assigning $\mu_s = 2,5,10~\mathrm{cm}^{-1}$ in the top layers, with deeper layer again held at $30~\mathrm{cm}^{-1}$. These values fall within physiologically realistic ranges with previously mentioned literature and semi-empirical models \citep{Graaff1992,Tuchin1997}. Accordingly, they were selected to isolate the qualitative effects of superficial scattering modulation rather than to reproduce exact experimental optical properties.

The absorbing ink layer was assigned concentration-dependent absorption coefficients spanning $\mu_a = 1$--$100~\mathrm{cm}^{-1}$ based on experimental calibration reported in the literature \citep{Ninni2010}, while all other optical parameters were held constant.

From the simulations, surface diffuse reflectance was recorded at $z = 0$ and radially binned as a function of source–detector separation to enable direct comparison with CW measurements. Depth-resolved photon fluence distributions were computed in the $x$–$z$ plane to visualize changes in photon penetration depth and sampling volume (“banana” profiles) under different scattering conditions. Photon interactions within the absorbing layer were quantified using the trapped-photon fraction $N_{\mathrm{i}}/N_{\mathrm{t}}$, where $N_{\mathrm{i}}$ denotes the number of photons interacting with the absorber and $N_{\mathrm{t}}$ is the total number launched. In addition, absorption contrast maps were computed from logarithmic ratios of photon fluence distributions with and without the absorber. Together, these metrics enable direct assessment of how superficial scattering reduction redistributes photon pathlengths and enhances interaction probability with subsurface absorbers in the measurements.

\section{Results and Discussion}
\subsection{Spectrometer measurements for Tartrazine-induced sample}
\begin{figure}[t]
  \centering
  \includegraphics[width=1\linewidth]{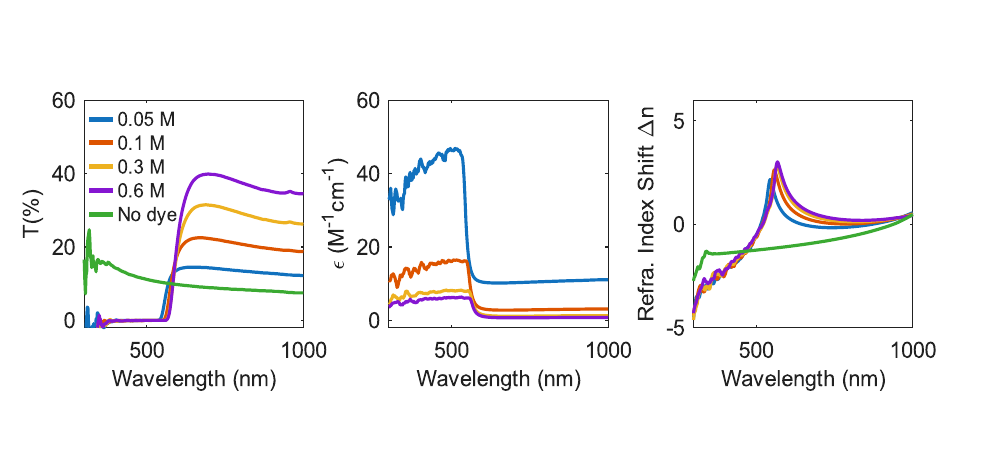}
  \caption{Optical property changes induced by Tartrazine in a silica particle suspension (1--5~$\mu$m) used as a controlled scattering medium. (a) Transmittance spectra, (b) molar absorption spectra, and (c) refractive-index shift as a function of wavelength for increasing Tartrazine concentrations. The systematic refractive-index increase across the visible--NIR range leads to reduced scattering despite absorption features at shorter wavelengths.}
  \label{fig:OpticalProperties_Combined}
\end{figure}

To establish the physical basis for Tartrazine-induced optical clearing used in the tissue experiments, we first examined how increasing dye concentration modifies optical properties in a controlled scattering medium. Measurements from the spectrometer in Fig.~\ref{fig:OpticalProperties_Combined} show the effect of increasing Tartrazine concentration on optical properties in a silica particle scattering sample. As the dye concentration increases, the measured transmittance increases across the visible–NIR range, indicating enhanced optical transparency due to reduced scattering. The molar absorption spectra confirm that Tartrazine exhibits strong absorption below approximately 600~nm, with absorption rapidly decreasing at longer wavelengths. Beyond this absorption band, the effective attenuation decreases with increasing dye concentration, consistent with reduced scattering dominating the measured transmittance.

The refractive-index change of the surrounding medium was estimated from the measured absorption spectra using Kramers–Kronig relations, yielding a relative, wavelength-dependent RI increase with increasing Tartrazine concentration across the visible–NIR range. Although this approach does not provide absolute RI, it robustly captures concentration-dependent trends. Together, these results confirm that the observed increase in transparency arises primarily from RI elevation of the interstitial medium rather than structural modification of the scattering particles, establishing RI matching as the physical mechanism underlying Tartrazine-induced scattering suppression in the red–NIR spectral region.

\subsection{Experimental results: chicken tissue vs uniform intralipid phantom}

\begin{figure}[t]
  \centering
  \includegraphics[width=1.0\linewidth]{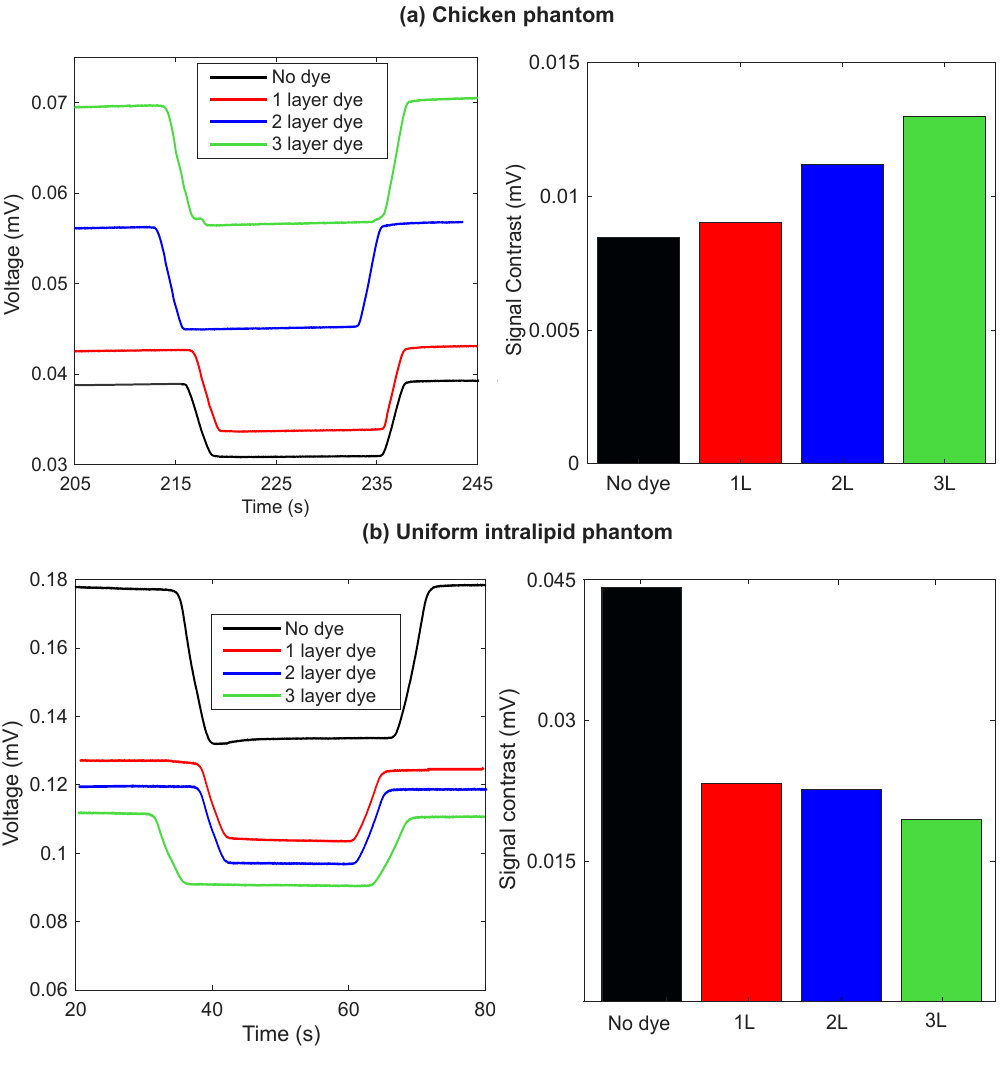}
 \caption{CW diffuse reflectance measurements at a 6 cm source--detector separation for (a) chicken tissue phantom and (b) uniform intralipid phantom under no-dye and increasing dye layer conditions. Left panels show time-domain voltage traces during ink infusion; right panels show the corresponding extracted signal contrast.}
  \label{fig:2phantomBar_line6cmplot}
\end{figure}

Figure~\ref{fig:2phantomBar_line6cmplot} presents CW diffuse reflectance measurements acquired at a 6~cm source–detector separation to experimentally evaluate how optical clearing–induced scattering modulation influences subsurface absorber detectability. By comparing a layered biological phantom with a homogeneous scattering phantom, this experiment isolates the role of depth-dependent versus uniform scattering reduction on CW signal contrast. Measurements were performed on (a) a layered chicken tissue phantom and (b) a homogeneous intralipid phantom under no-dye and increasing Tartrazine layer conditions.

For both phantoms, the detected voltage remains stable prior to ink injection, exhibits a sharp decrease during ink infusion, and recovers after ink withdrawal, demonstrating reproducible and time-stable CW measurements. The voltage difference between the baseline (no ink) and ink-attenuated states is used to quantify signal contrast. In the chicken tissue phantom (Fig.~\ref{fig:2phantomBar_line6cmplot}a), increasing the number of Tartrazine layers produces a systematic increase in baseline diffuse reflectance intensity and a monotonic enhancement in ink-induced signal contrast for a fixed ink concentration (1\% v/v). This trend is evident in both the time-domain voltage traces (left) and the extracted signal contrast values (right), indicating improved detectability of the subsurface absorber at large source–detector separation under optically cleared, depth-dependent scattering conditions.

In contrast, the uniform intralipid phantom (Fig.~\ref{fig:2phantomBar_line6cmplot}b) exhibits the opposite trend. Here, dye addition leads to a monotonic reduction in both baseline intensity and signal contrast. Because the intralipid phantom undergoes a uniform scattering reduction throughout its volume, photons become more forward-directed, reducing backscattered diffuse reflectance at the surface. In the absence of a depth-dependent scattering gradient, optical clearing does not enhance sensitivity to the subsurface absorber.

\subsection{Model predictions for diffuse reflectance in tissue phantom}

The experimental results suggest that optical clearing modifies diffuse reflectance primarily by altering photon transport near the tissue surface rather than uniformly throughout the volume. To interpret this behavior and isolate the underlying mechanism, we examined how superficial scattering reduction redistributes diffuse reflectance using complementary analytical and numerical models.
\begin{figure}[t]
  \centering
  \includegraphics[width=1\linewidth]{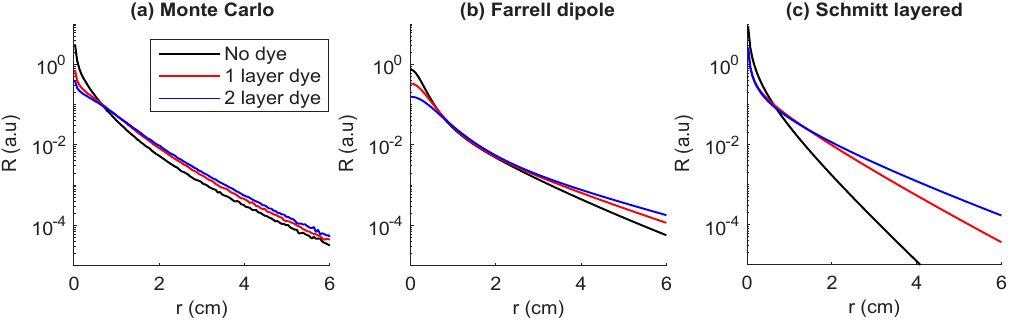}
  \caption{Radial dependence of diffuse reflectance predicted by Monte Carlo simulations for layered tissue phantom, the Farrell steady-state diffusion dipole model, and the Schmitt layered diffusion model for no-dye, one-layer dye, and two-layer dye conditions. All three approaches show reduced reflectance near the source and enhanced reflectance at larger radial distances as superficial scattering is reduced.}
  \label{fig:Model_combined}
\end{figure}
Figure~\ref{fig:Model_combined} compares the radial dependence of diffuse reflectance predicted by Monte Carlo simulations and two diffusion-based models under conditions representing no clearing, moderate, and stronger superficial clearing in layered chicken tissue phantoms. In the diffusion-based models, scattering reduction is incorporated through a decrease in the effective attenuation coefficient $\mu_{\mathrm{eff}}$, either uniformly (Farrell model) or within designated superficial layers (Schmitt model). In the Monte Carlo simulations, the same effect is implemented by explicitly assigning lower scattering coefficients to the superficial layers while preserving deeper-layer properties. Although the models differ in formulation, all are designed to test the same physical hypothesis that reducing superficial scattering alters the spatial distribution of photon escape.

Across all three approaches, a consistent qualitative trend is observed. At short source–detector separations ($r \lesssim 0.5~\mathrm{cm}$), optically cleared conditions exhibit reduced reflectance relative to the no-dye case, indicating suppression of early backscattering from superficial tissue. As radial distance increases, a crossover behavior emerges, beyond which reflectance from the cleared cases exceeds that of untreated tissue. This enhancement becomes more pronounced as the extent of superficial scattering reduction increases, with the two-layer clearing condition producing the highest reflectance at large separations.

These results indicate that optical clearing redistributes diffuse reflectance radially rather than uniformly increasing signal levels. Reduced superficial scattering decreases the probability of early photon escape near the source and increases the relative contribution of longer photon paths that re-emerge at larger lateral distances. The agreement among Monte Carlo simulations and both diffusion-based models demonstrates that this redistribution is a robust physical consequence of superficial scattering reduction, rather than a model-specific artifact. Together, these predictions provide a mechanistic explanation for the experimentally observed enhancement in deep-signal detectability under optically cleared conditions.

\subsection{Weak-absorber detectability in tissue phantom}

The preceding experimental and modeling results demonstrate that optical clearing redistributes diffuse reflectance in layered tissue phantoms. We now test the central implication of this redistribution: whether it improves the detectability of weak subsurface absorbers in CW measurements under experimentally relevant conditions.

\begin{figure}[t]
  \centering
  \includegraphics[width=1\linewidth]{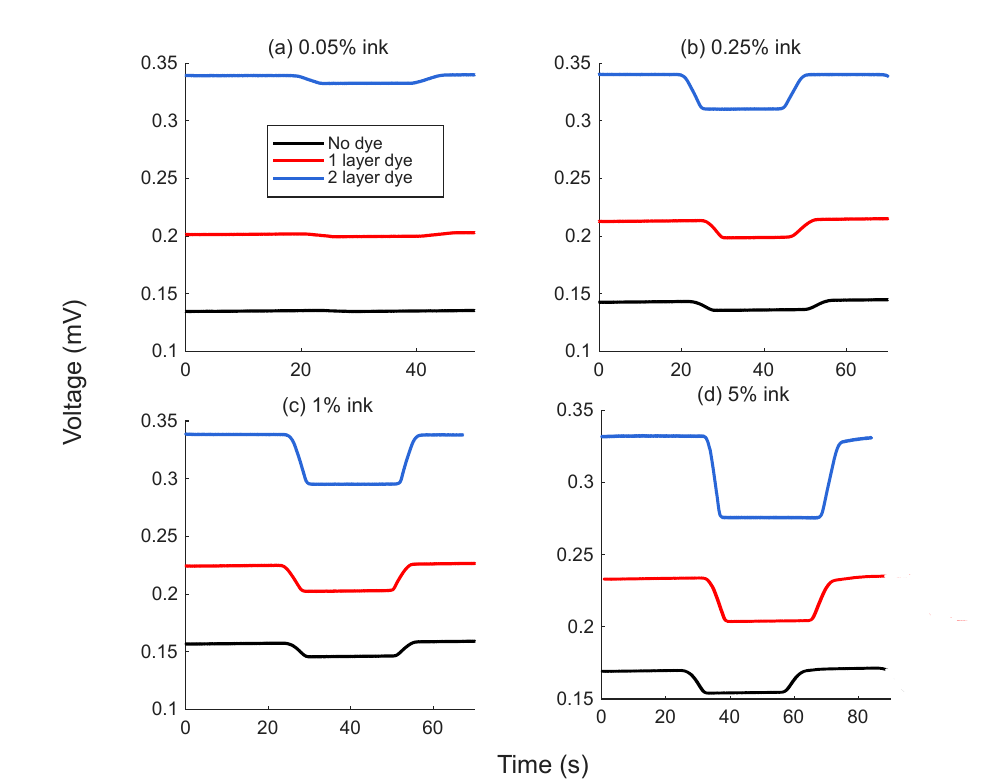}
  \caption{CW diffuse reflectance signals measured at a 6~cm source–detector separation for increasing ink strengths: (a) 0.05\%, (b) 0.25\%, (c) 1\%, and (d) 5\% v/v, under no-dye, one, and two-layer dye conditions. For weak absorbers (0.05\%), signal changes are indistinguishable from baseline fluctuations without optical clearing but become detectable after dye application. At higher ink concentrations, signal contrast increases for all conditions, with consistently larger contrast observed under optically cleared states.}
  \label{fig:AllinkAlldye6cm}
\end{figure}

Figure~\ref{fig:AllinkAlldye6cm} shows CW diffuse reflectance signals measured at a fixed source–detector separation of 6~cm for increasing ink concentrations under no-dye, one-layer dye, and two-layer dye conditions. For the weakest absorber (0.05\% v/v) shown in Fig.~\ref{fig:AllinkAlldye6cm}(a), the ink-induced signal change is not distinguishable from baseline fluctuations in the absence of optical clearing. After applying one dye layer, the signal contrast increases by approximately threefold, and with two dye layers by about five- to sixfold, producing a clear and reproducible response. This demonstrates that optical clearing enables detection of an otherwise undetectable weak absorber at depth.

As ink concentration increases (Fig.~\ref{fig:AllinkAlldye6cm}(b--d)), absorption-induced signal changes become apparent for all conditions. However, at each concentration level, optically cleared states consistently yield larger signal contrast than the no-dye condition. The magnitude of this enhancement increases with the degree of superficial scattering reduction, with two-layer clearing producing the strongest contrast.

These results provide direct experimental evidence that optical clearing improves weak-absorber detectability in CW diffuse reflectance measurements. Importantly, the enhancement arises not from an increase in overall detected intensity, but from a redistribution of the detected photon population toward longer pathlengths that more effectively sample the subsurface absorbing region.

\subsection{Monte Carlo simulation results}

\begin{figure}[t]
  \centering
  \includegraphics[width=1\linewidth]{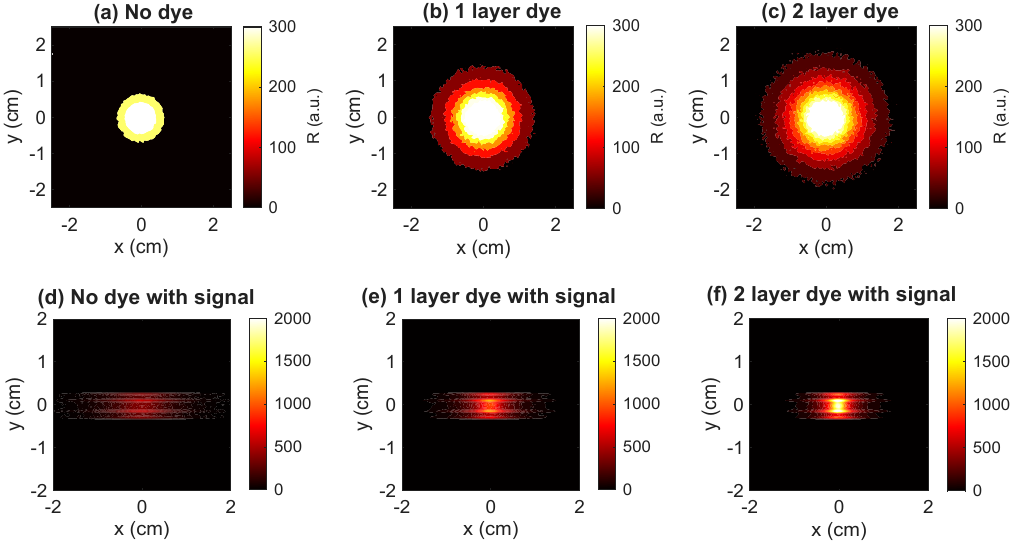}
  \caption{Monte Carlo simulation results illustrating photon redistribution induced by optical clearing. (a–c) Surface diffuse reflectance maps ($z = 0$) for no-dye, one-layer dye, and two-layer dye conditions, showing lateral broadening and increased reflectance at larger distances with reduced superficial scattering. (d–f) Spatial distributions of photons interacting with the ink layer, along with the corresponding trapped-photon fraction ($N_{\mathrm{i}}/N_{\mathrm{t}}$).}
  \label{fig:Combined_Surface_diff_refl_Ink_XY_namedef}
\end{figure}

Monte Carlo simulations were used to examine how superficial scattering reduction alters photon propagation, surface reflectance, and interaction with a subsurface absorber. Figures~\ref{fig:Combined_Surface_diff_refl_Ink_XY_namedef}(a--c) show surface diffuse reflectance distributions ($z = 0$) for the no-dye, one-layer dye, and two-layer dye conditions. In the no-dye case, reflected photons are concentrated near the source, with reflectance decaying rapidly with lateral distance. As superficial scattering is reduced, the reflectance distribution broadens laterally, with increased signal observed at larger radial distances from the source.

To quantify how this redistribution affects subsurface sampling, Figs.~\ref{fig:Combined_Surface_diff_refl_Ink_XY_namedef}(d--f) show the spatial distribution and fraction of photons interacting with the ink layer, expressed as the trapped-photon ratio $N_{\mathrm{i}}/N_{\mathrm{t}}$, where $N_{\mathrm{i}}$ denotes the number of photons interacting with the absorber and $N_{\mathrm{t}}$ is the total number of launched photons. This fraction increases monotonically with increasing scattering reduction, from $1.36\times10^{-1}$ (no dye) to $1.96\times10^{-1}$ (one layer) and $2.40\times10^{-1}$ (two layers), indicating an increased probability of photon–absorber interaction under optically cleared conditions. In addition, increasing dye layers produce a more spatially concentrated photon distribution within the absorber region, reflecting reduced lateral diffusion and more directed photon delivery to depth.

\begin{figure}[t]
  \centering
  \includegraphics[width=1\linewidth]{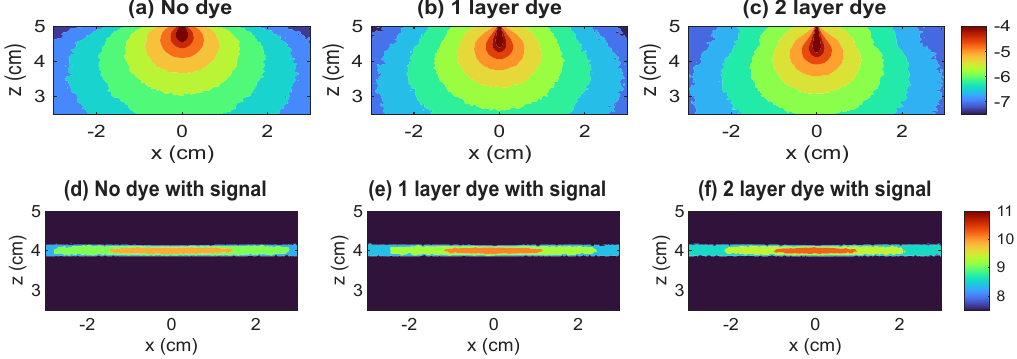}
  \caption{Depth-resolved photon transport and absorption contrast obtained from Monte Carlo simulations. (a–c) Banana-shaped photon intensity distributions in the absence of the ink absorber for no-dye, one-layer dye, and two-layer dye conditions, showing increased penetration depth as superficial scattering is reduced. (d–f) Absorption contrast maps computed from the logarithmic ratio of photon intensity distributions without and with the ink absorber.}
  \label{fig:Combined_Banana_InkAbsorbanceXZ_3dyeconditions}
\end{figure}

Figures~\ref{fig:Combined_Banana_InkAbsorbanceXZ_3dyeconditions}(a--c) show depth-resolved photon fluence distributions in the absence of the ink absorber. In the no-dye condition, photon density is concentrated near the surface, indicating limited penetration toward the absorber depth. As superficial scattering is reduced, the photon sampling volume extends deeper, resulting in increased fluence at the depth of the ink layer.

The corresponding absorption contrast maps in Figs.~\ref{fig:Combined_Banana_InkAbsorbanceXZ_3dyeconditions}(d--f), computed from logarithmic fluence ratios with and without the absorber, reveal a clear enhancement of absorption signatures with increasing optical clearing. For a fixed ink absorption coefficient ($\mu_a = 100~\mathrm{cm^{-1}}$), optical clearing produces stronger and more spatially localized contrast at depth, consistent with increased photon delivery to the subsurface absorber rather than a change in absorption strength.

\begin{figure}[t] \centering \includegraphics[width=1\linewidth]{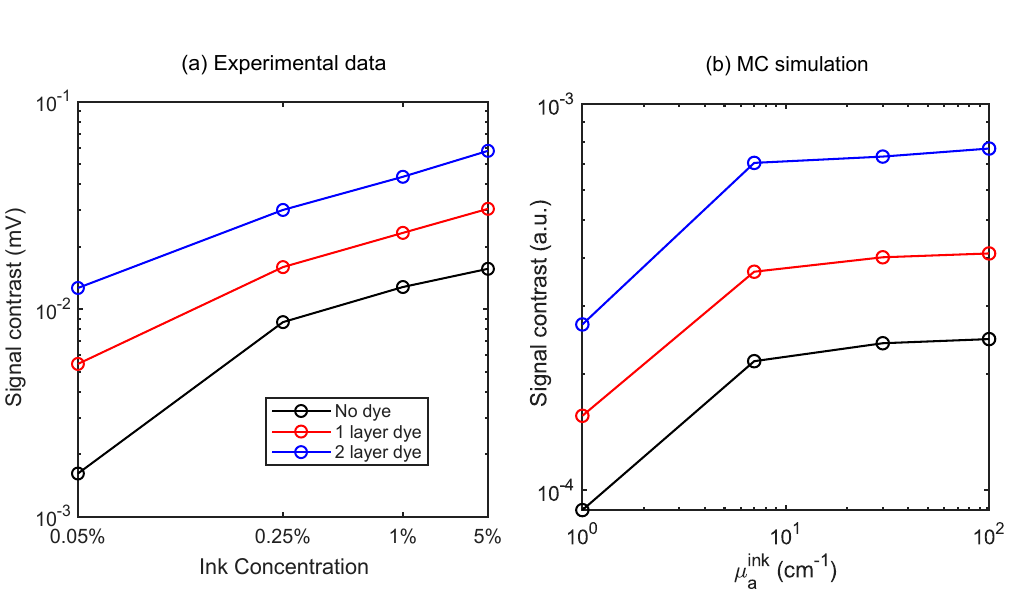} \caption{Comparison of signal contrast as a function of ink concentration for (a) experimental CW diffuse reflectance measurements and (b) Monte Carlo simulations at a fixed source–detector separation of 6 cm. In both datasets, signal contrast increases monotonically with ink concentration, and dye-treated conditions consistently yield larger contrast than the no-dye case. The preserved relative ordering among dye conditions confirms that superficial scattering reduction is the dominant mechanism governing deep-signal enhancement.} \label{fig:Combined_signalcontrast_Exp_MC_pos} \end{figure}
The combined experimental, analytical, and Monte Carlo results consistently demonstrate that optical clearing enhances deep-tissue signal detectability by altering the statistical distribution of photon trajectories contributing to the detected signal. Reducing superficial scattering suppresses early backscattering and increases the relative contribution of longer pathlength photons that sample deeper tissue regions. This redistribution of the detected photon ensemble increases the probability of interaction with subsurface absorbers, leading to enhanced absorption contrast without changing the intrinsic absorption strength or introducing optical gain.

This mechanism is directly reflected in the quantitative agreement between experiment and simulation shown in Fig.~\ref{fig:Combined_signalcontrast_Exp_MC_pos}. For both measured CW diffuse reflectance data and Monte Carlo simulations at a fixed 6~cm source–detector separation, signal contrast increases monotonically with ink concentration, and optically cleared conditions consistently yield higher contrast than the no-dye case. Importantly, the preserved ordering among no-dye, one-layer dye, and two-layer dye conditions across all absorber strengths confirms that superficial scattering reduction—rather than changes in absorption or total photon throughput—is the dominant factor governing deep-signal enhancement in CW diffuse reflectance measurements.

\section{Conclusion}

This study demonstrates that optical scatter reduction significantly improves the detectability of weak absorbers at depth in continuous-wave (CW) diffuse reflectance measurements. The improvement in reflectance arises from enhanced photon penetration and increased absorption at depth for absorbers of identical strength as well as the gradient in refractive index created as the dye penetrates the tissue.

Using controlled phantom experiments, analytical diffusion-based models, and Monte Carlo simulations, we show that reducing tissue scattering suppresses early backscattering near the surface and allows a greater fraction of photons to propagate deeper into the tissue. This redistribution leads to larger measurable intensity changes at the detector, particularly at large source--detector separations where CW measurements are typically limited.

Experimental results demonstrate a consistent increase in signal contrast with increasing scatter reduction for a fixed absorber strength. Diffusion models predict a corresponding redistribution of diffuse reflectance toward larger radial distances, and Monte Carlo simulations confirm that optical clearing increases the probability that photons reach and interact with subsurface absorbers without altering their absorption properties.

While the present work focuses on CW measurements, the underlying physical mechanism, reduced superficial scattering leading to deeper photon penetration and increased absorption at depth, is general. Similar scattering reduction is expected to benefit frequency-domain and time-domain diffuse optical systems by improving photon penetration and potentially enhancing quantitative accuracy, which could be validated from further experiments. Our experiments also show the need for developing better optical phantoms that can mimic the gradient effect as the dye diffuses in real tissue. 

Although the present study was conducted using ex vivo tissue phantoms, the identified transport mechanism is applicable to other highly scattering soft tissues, including breast tissue. Together, these findings establish tartrazine like dye-induced scatter reduction as a physically grounded and practical strategy for improving depth sensitivity in CW diffuse reflectance imaging.

\subsection* {Funding} 

National Institute of Biomedical Imaging and Bioengineering (R01 EB029761); Congressionally Directed Medical Research Programs (BC151607); National Science Foundation (1652892).

\subsection* {Acknowledgments}
The authors acknowledge the use of the Carya Cluster and the advanced support
provided by the Research Computing Data Core at the University of Houston in carrying out the research
presented in this study.

\subsection*{Disclosures}
 The authors declare no conflicts of interest.

\subsection* {Code, Data, and Materials Availability} 
Data underlying the results presented in this study are not publicly available at this time
but may be obtained from the authors upon reasonable request.

\listoffigures
\listoftables
\bibliographystyle{spiebib}
\bibliography{report}

\end{spacing}
\end{document}